\documentclass[a4paper,11pt]{article}
\usepackage{pos}

\usepackage{bm}
\usepackage{ulem}

\newcommand{\gtapprox}{\raisebox{-0.5ex}{$\,\stackrel{>}{\scriptstyle\sim}\,$}}
\newcommand{\ltapprox}{\raisebox{-0.5ex}{$\,\stackrel{<}{\scriptstyle\sim}\,$}}

\title{Computing $1/m_Q$ and $1/m_Q^2$ corrections to the static potential with lattice gauge theory using gradient flow}
\ShortTitle{Computing $1/m_Q$ and $1/m_Q^2$ corrections to the static potential}

\author*[a,b]{Michael Eichberg}
\author[a,b]{Marc Wagner}

\affiliation[a]{Goethe-Universit\"at Frankfurt am Main, Institut f\"ur Theoretische Physik, Max-von-Laue-Stra{\ss}e 1, D-60438 Frankfurt am Main, Germany}

\affiliation[b]{Helmholtz Research Academy Hesse for FAIR, Campus Riedberg, Max-von-Laue-Stra{\ss}e 12, \\ D-60438 Frankfurt am Main, Germany}

\emailAdd{eichberg@itp.uni-frankfurt.de}
\emailAdd{mwagner@itp.uni-frankfurt.de}

\abstract{We present selected preliminary lattice gauge theory results for $O(1/m_Q)$ and $O(1/m_Q^2)$ corrections to the static potential. These results are based on Wilson loops with two field strength insertions, which we renormalize using gradient flow. We explore tree level improvement to reduce systematic errors in the Wilson loops due to the finite lattice spacing and flow time, in particular at small temporal and spatial separations.
}

\FullConference{The 41st International Symposium on Lattice Field Theory (LATTICE2024)\\
 28 July - 3 August 2024\\
Liverpool, UK\\}

\begin{document}
\maketitle


\section{Introduction}

There are numerous approaches to study quarkonium, which use the static quark-anti-quark potential as well as corrections proportional to powers of $1/m_Q$ \cite{Eichten:1980mw,Barchielli:1986zs}, where $m_Q$ is the heavy quark mass. 
Potential Non-Relativistic QCD (pNRQCD) \cite{Brambilla:1999xf,Brambilla:2000gk,Pineda:2000sz}, which is an effective theory, is a prominent example.
Using these potentials in a Schr\"odinger equation allows to compute spectra and properties of heavy quarkonium systems (see e.g.\ Refs.\ \cite{Brambilla:2004jw,Berwein:2024ztx}). 

Corrections due to the finite heavy quark mass have been calculated up to N$^3$LO and N$^3$LL accuracy in perturbation theory \cite{Anzai:2018eua,Peset:2018jkf}. 
In lattice QCD these potentials have first been studied in Refs.\ \cite{deForcrand:1985zc,Campostrini:1986ki} and later in Refs.\ \cite{Bali:1997am,Bali:1998pi}.
A technical problem in such computations are the rather large statistical fluctuations of the field strength correlators.
Using the multilevel algorithm Ref.\ \cite{Koma:2006fw} achieved unprecedented numerical precision, but with open questions related to renormalization and lattice discretization errors remaining.
In this work we try to overcome these problems using gradient flow, continuing previous work reported in Ref.\ \cite{Eichberg:2023trq}.


\section{The heavy quark-anti-quark potential up to $\mathcal{O}(1/m_Q^{2})$}

\begin{figure}[ht]
\centering
\includegraphics[width=0.3\linewidth]{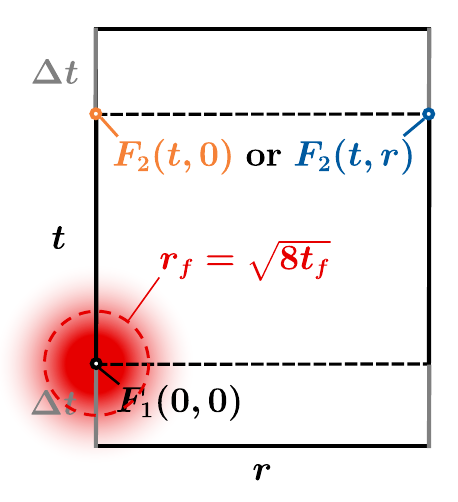}
\caption{Generalized Wilson loop with two field strength insertions $W_{r \times (t + 2 \Delta t)}(F_2(t, r_2), F_1(0, 0))$. The flow radius $r_f = \sqrt{8 t_f}$ is indicated by a red circle.}
\label{fig:loop_sketch}
\end{figure}

The potential describing a heavy quark-anti-quark pair for quark masses $m_{Q_1} = m_{Q_2} = m_Q$ is
\begin{align}
V
&=
V^{(0)} + \frac{1}{m_Q} V^{(1)} + \frac{1}{m_Q^2} \left( V_\mathrm{SD}^{(2)} + V_\mathrm{SI}^{(2)} \right) + \mathcal{O}(1/m_Q^3)
\label{eq:potential}
\end{align}
\cite{Pineda:2000sz}.
It is composed of the static potential $V^{(0)}$, a $1/m_Q$ correction $V^{(1)}$ and spin dependent (SD) as well as spin independent (SI) $1/m_Q^2$ corrections $V_\mathrm{SD}^{(2)}$ and $V_\mathrm{SI}^{(2)}$. 
$V^{(1)}$, $V_\mathrm{SD}^{(2)}$ and $V_\mathrm{SI}^{(2)}$ can be computed by solving integrals of the type
\begin{align}
\int_0^\infty \mathrm{d}t\, t^s \langle \Sigma_g^+, r \vert F_2(t, r_2) F_1(0, 0) \vert \Sigma_g^+, r\rangle\,, \quad s=0, 1, 2\,,\quad r_2 = 0, r\,,
\label{eq:integral_type}
\end{align}
where $F_{1,2} = B_i, E_i$ ($i=1,2,3$) are chromo-magnetic and chromo-electric fields and $\vert \Sigma_g^+, r\rangle$ denotes the ground state of a static quark-anti-quark pair with separation $r$ connected by a flux tube with quantum numbers $\Sigma_g^+$.
The correlator in Eq.\ (\ref{eq:integral_type}) can be expressed in terms of a generalized Wilson loops with two field strength insertions,
\begin{align}
\langle \Sigma_g^+, r \vert F_2(r_2, t) F_1(0, 0) \vert \Sigma_g^+, r\rangle
&=
\lim_{\Delta t \rightarrow \infty} \frac{W_{r \times (t + 2 \Delta t)}(F_2(t, r_2), F_1(0, 0))}{W_{r \times (t + 2 \Delta t)}}\,.
\end{align}
Fig.\ \ref{fig:loop_sketch} shows this generalized Wilson loop $W_{r \times (t + 2 \Delta t)}(F_2(t, r_2), F_1(0, 0))$.
For the field strength insertions we use the clover definition
\begin{align}
B_i = \frac{\epsilon_{ijk}}{2i} \left( \Pi_{jk} - \Pi_{jk}^\dagger \right),\ E_i = \frac{1}{2i} \left( \Pi_{i0} - \Pi_{i0}^\dagger \right),\ \Pi_{\mu\nu} = \frac{1}{4} \left( P_{\mu, \nu} + P_{\nu, -\mu} + P_{-\mu, -\nu} + P_{-\nu, \mu} \right)\,,
\end{align}
where $P_{\mu, \nu}$ denotes the plaquette.

\section{\label{SEC_gradient_flow}Gradient flow, renormalization and tree level improvement}

When computing potential corrections, one has to face the following problems: 1.\ Generalized Wilson loops with field strength insertions exhibit poor signal-to-noise ratios. 2.\ Field strength insertions require renormalization. 3.\ Chromo-magnetic fields are logarithmically divergent proportional to $1/a$ and, thus, field strength correlators have no continuum limit.
In principle, these problems can be overcome using gradient flow \cite{Luscher:2010iy}.
Evolving the gauge field to flow time $t_f > 0$ using the flow equation
\begin{align}
\dot{B}_\mu
&=
D_\nu G_{\mu\nu}\,,
\end{align}
where
\begin{align}
\label{EQN_gradient_flow} B_\mu \Big\vert_{t_f=0} = A_\mu,\quad 
G_{\mu\nu} = \partial_\mu B_\nu - \partial_\nu B_\mu + [B_\mu, B_\nu],\quad
D_\mu = \partial_\mu + [B_\mu, \cdot]\,,
\end{align}
one can show that correlation functions require no additional renormalization \cite{Luscher:2011bx} (see also the recent study of generalized Wilson loops with just one chromo-electric field insertion \cite{Brambilla:2023fsi}). UV-fluctuations are suppressed, which leads to significantly enhanced signal-to-noise ratios.
Moreover, divergencies in spin-dependent potential corrections are regulated by $t_f$ and results can be extrapolated to the continuum. 
NRQCD matching coefficients, which are necessary to relate the spin-dependent potential corrections in the gradient flow scheme to the $\overline{\mathrm{MS}}$ scheme, have been computed up to NLO in Ref.\ \cite{Brambilla:2023vwm}. In contrast to that, spin-independent potential corrections do not require matching coefficients and can be obtained by performing a combined $a\rightarrow 0$, $t_f\rightarrow 0$ extrapolation, for the latter typically using a small flow time expansion.

Note that gradient flow also introduces systematic errors in correlators with small separations $d$ between operators, in particular the field strength insertions $F_1$ and $F_2$ (see Fig.\ \ref{fig:loop_sketch}). 
These unwanted flow effects are strongly suppressed for $d \gtapprox 2 r_f = 2 \sqrt{8 t_f}$ \cite{Eller:2018yje}, where $r_f$ is the flow radius.
In Ref.\ \cite{Brambilla:2023fsi} flow effects in the static force are reduced using tree level improvement, 
by multiplying with the ratio of the continuum and lattice tree level expressions of the static force at finite $t_f$.
Here we use a similar approach for the static potential, but instead of a multiplicative factor we parametrize tree level flow effects by a correction term \cite{Bali:1997am,Schlosser:2021wnr}. 
We carry out a fit to lattice data points for $V^{(0)}$ using the ansatz
\begin{align}
V^{(0)}(\bm{r}, t_f)
&=
- \frac{c^{(0)}}{r} + \sigma r + V_c(t_f, a) + \tilde{c}^{(0)} \Big( 4\pi G(\bm{r}, t_f) - 1/r \Big) \, .
\label{eq:ansatz_tree_level}
\end{align}
The first two terms are equivalent to the Cornell ansatz. 
$V_c(t_f, a)$ is an $r$-independent shift, which depends, however, on the lattice spacing $a$ and the flow time $t_f$. 
The last term is proportional to the difference between the tree level results in lattice and continuum perturbation theory. 
For lattice gauge link configurations generated with the Wilson plaquette action and evolved to flow time $t_f$ via the prescription (\ref{EQN_gradient_flow}), $G(\bm{r}, t_f)$ is given by
\begin{align}
G(\bm{r}, t_f)
&=
\int_{[-\pi,\pi)^3} \frac{\mathrm{d}^3p}{(2\pi)^3} e^{i\bm{p}\bm{r}} \frac{e^{- 2 t_f \hat{\bm{p}}^2}}{\hat{\bm{p}}^2}\,,\quad \hat{p}_i = 2 \sin(p_i/2)
\end{align}
\cite{Brambilla:2023fsi}.
Extracting the parameters $c^{(0)}$ and $\sigma$ (as well as $V_c(t_f, a)$ and $\tilde{c}^{(0)}$) from a global fit of the ansatz (\ref{eq:ansatz_tree_level}) to data points computed at different lattice spacings and flow times results in an $a\rightarrow 0$ and $t_f\rightarrow 0$ extrapolated Cornell parameterization of the static potential.
A similar strategy for tree level improving field strength correlators is outlined in Sec.\ \ref{SEC_results_corrections}.


\section{Numerical results}


\subsection{Lattice setup}

Table \ref{tab:ensembles} provides details on the SU(3) gauge link ensembles used in this work. 
Correlators have been computed at flow radii $r_f = \sqrt{8t_f} \approx 0.088, 0.102, 0.111, 0.119$ fm. 
Errors have been propagated using \texttt{\href{https://fjosw.github.io/pyerrors/pyerrors.html}{pyerrors}} \cite{Joswig:2022qfe} and the Gamma method \cite{Wolff:2003sm} and automatic differentiation has been used for fitting \cite{Ramos:2018vgu}.

\begin{table}[ht]
\centering
\begin{tabular}{c|c|c|c|c}
ensemble & $\beta$ & $(T/a)\cdot (L/a)^3$ & $a$ [fm] & $N_\mathrm{conf}$
\\
\hline
A & $6.284$ & $48\cdot 24^3$ & $0.06\phantom{0}$ & $10000$
\\
\hline
B & $6.451$ & $60\cdot 30^3$ & $0.048$ & $\phantom{0}4800$
\\
\hline
C & $6.594$ & $72\cdot 36^3$ & $0.04\phantom{0}$ & $\phantom{0}2400$
\end{tabular}
\caption{Gauge link ensembles generated using \texttt{\href{https://zenodo.org/records/5121917}{CL2QCD}} \cite{cl2qcd}. Physical units were introduced by setting $r_0 = 0.5 \, \text{fm}$.}
\label{tab:ensembles}
\end{table}


\subsection{\label{SEC_results_V0}The ordinary static potential}

\begin{figure}[ht]
\centering
\includegraphics[width=0.48\textwidth]{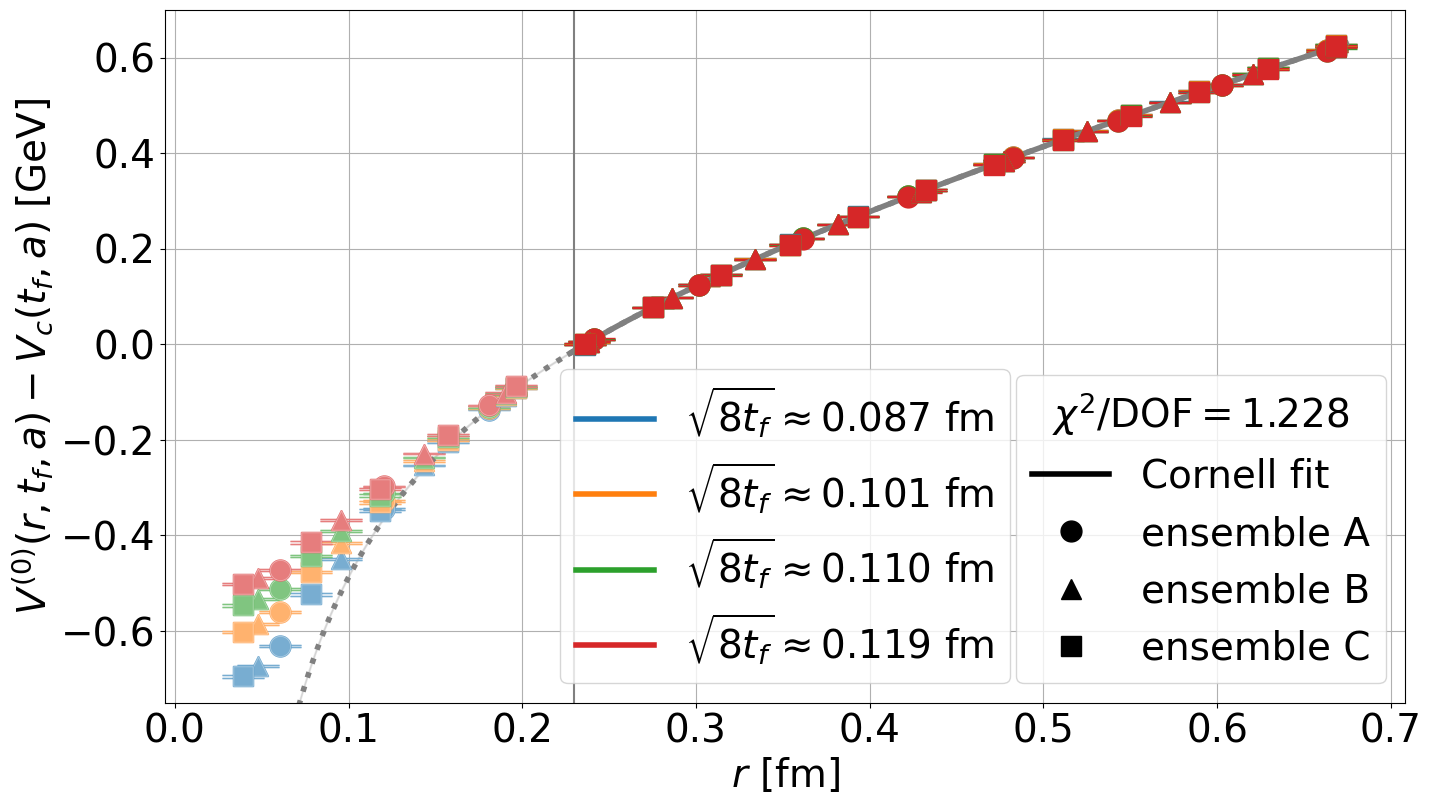}
\includegraphics[width=0.48\textwidth]{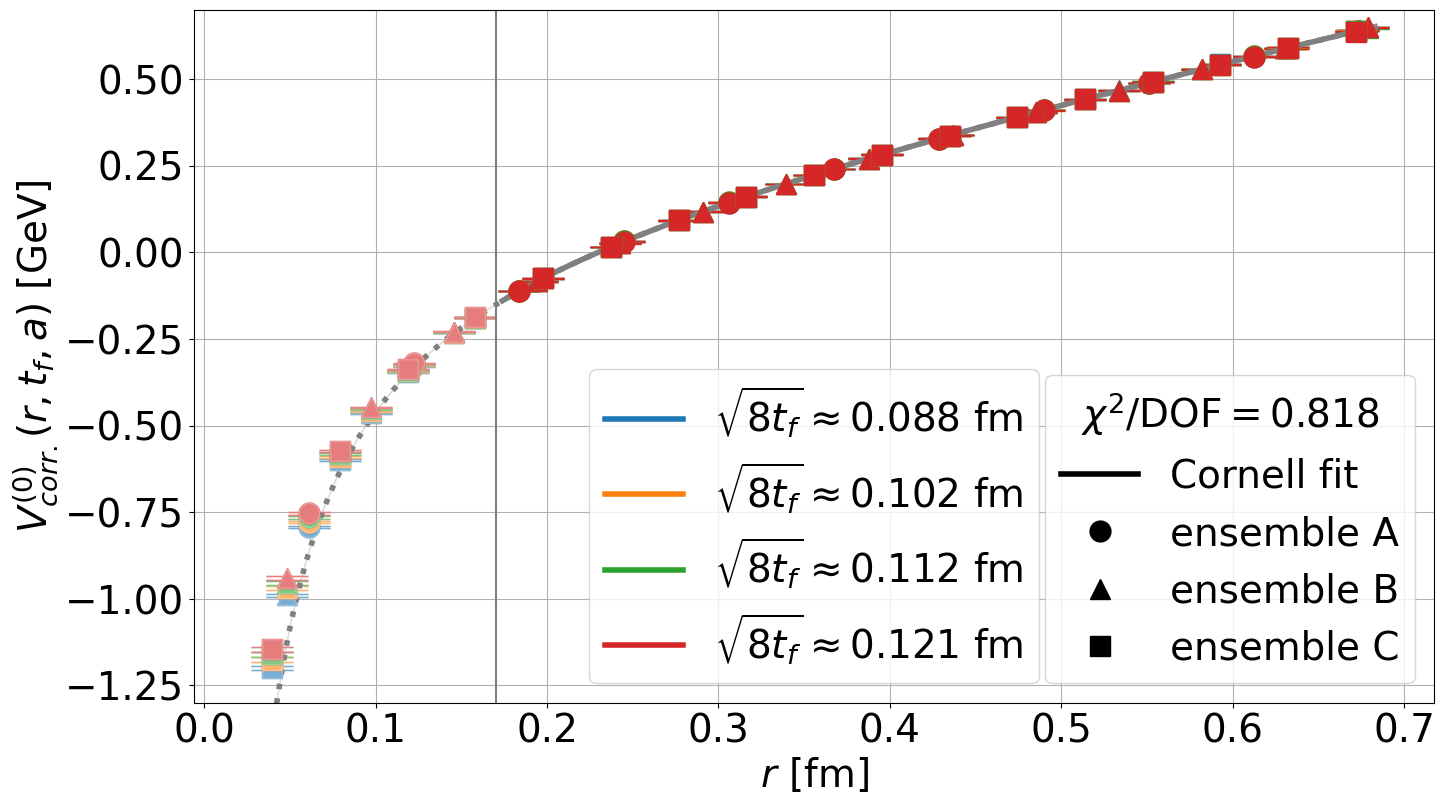}
\caption{Left: Static potential data points with mass shift $V_c(a, t_f)$ subtracted, as obtained by a fit with the ansatz (\ref{eq:ansatz_tree_level}) with fixed $\tilde{c}^{(0)} = 0$ (the vertical grey line indicates the lower bound of the fitting range). Flow effects are very prominent for $r \ltapprox 2 r_{f,\text{max}} \approx 0.24$ fm. Right: Tree level corrected potential data points, i.e.\ both $V_c(a, t_f)$ and $\tilde{c}^{(0)}(4\pi G(\bm{r}, t_f) - 1/r)$ have been subtracted. Flow effects are drastically reduced, allowing to include data points at smaller $r$ in the fit, neither increasing $\chi^2/\mathrm{DOF}$, nor changing the results for $c^{(0)}$ and $\sigma$ within errors.}
\label{fig:static_potential}
\end{figure}

Fig.\ \ref{fig:static_potential} shows results for the ordinary static potential $V^{(0)}$, i.e.\  the heavy quark-anti-quark potential without corrections due to the finite heavy quark mass. 
At small separations results strongly depend on the flow time (see left plot). For $r \gtapprox 2 r_{f,\text{max}} \approx 0.24 \, \mathrm{fm}$ this dependence is essentially gone.
After performing a common fit with the ansatz (\ref{eq:ansatz_tree_level}) to the data points obtained on all three ensembles and with each of the four flow times, we subtracted both the mass shift $V_c(t_f, a)$ and the tree level correction term proportional to $\tilde{c}^{(0)}$. This leads to data points for the static potential, which are in excellent agreement also at rather small separations $r \approx 0.05 \, \text{fm}$ (see right plot).
Our strategy for tree level improvement, thus, seems to be a successful method for a combined continuum and zero-flow-time extrapolation of the static potential.

\subsection{Corrections to the static potential due to the finite heavy quark mass}

In the following we do not discuss the complete $1/m_Q$ und $1/m_Q^2$ corrections appearing in Eq.\ (\ref{eq:potential}), but focus on four exemplary contributions:
$V_{LS}^{(2,0)}$, $V_{LS}^{(1,1)}$ and $V_{p^2}^{(1,1)}$ are part of the spin-dependent term $V_\mathrm{SD}^{(2)}$, while $V_{S_{12}}^{(1,1)}$ is part of the spin-independent term $V_\mathrm{SI}^{(2)}$.
Our strategy to solve the corresponding integrals (\ref{eq:integral_type}) is based on fits to lattice data points for the relevant field strength correlators and has been outlined in Ref.\ \cite{Eichberg:2023trq}.
The extraction of potential corrections from all three ensembles and the four used flow times is still ongoing. Tree level improvement, which is expected to be crucial for controlled continuum and $t_f\rightarrow 0$ extrapolations, has not been incorporated yet and spin-dependent potentials have not yet been converted from the gradient flow scheme to the $\overline{\mathrm{MS}}$ scheme.

\begin{figure}[ht]
\centering
\includegraphics[width=0.48\linewidth]{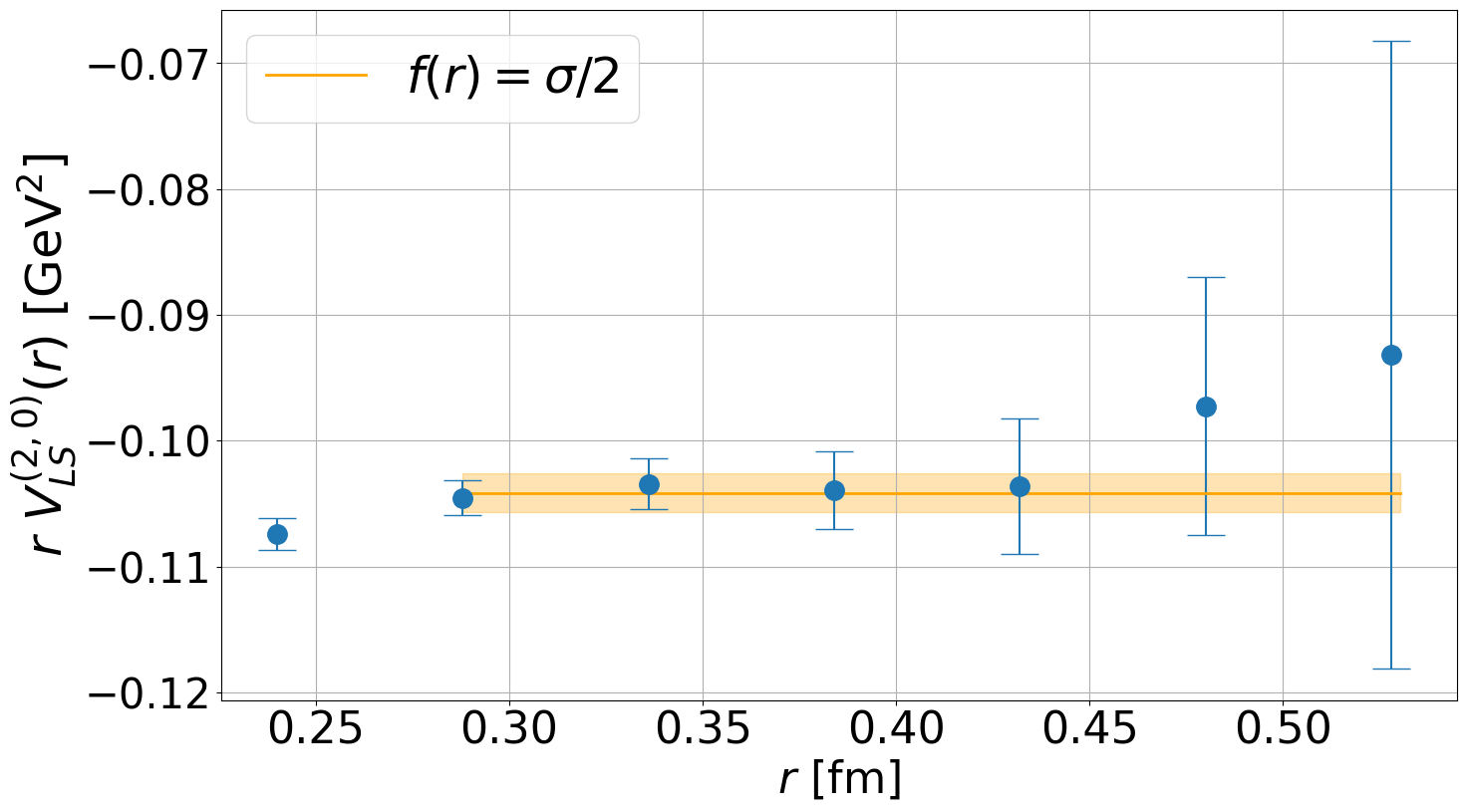}
\includegraphics[width=0.48\linewidth]{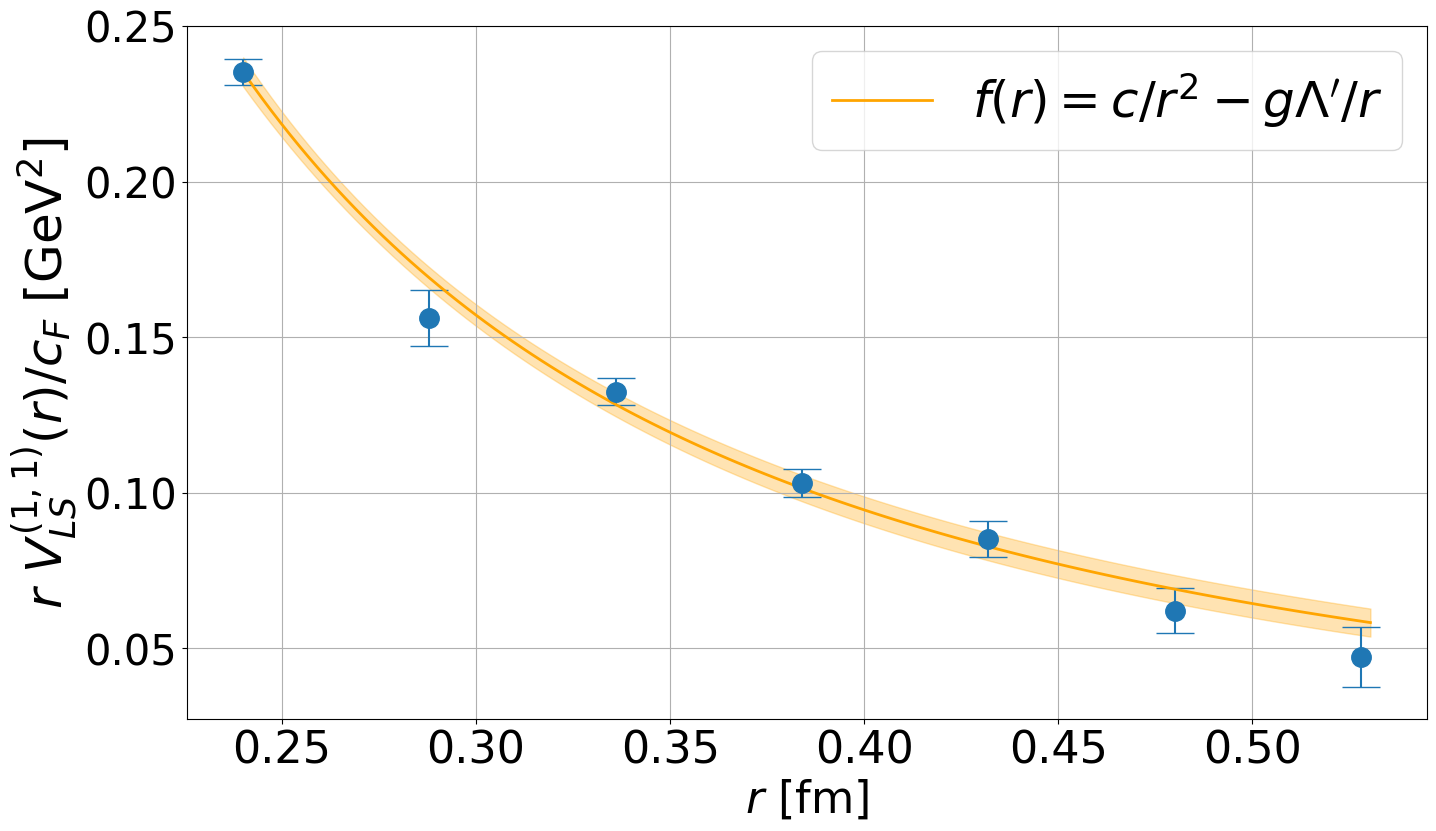}
\includegraphics[width=0.48\linewidth]{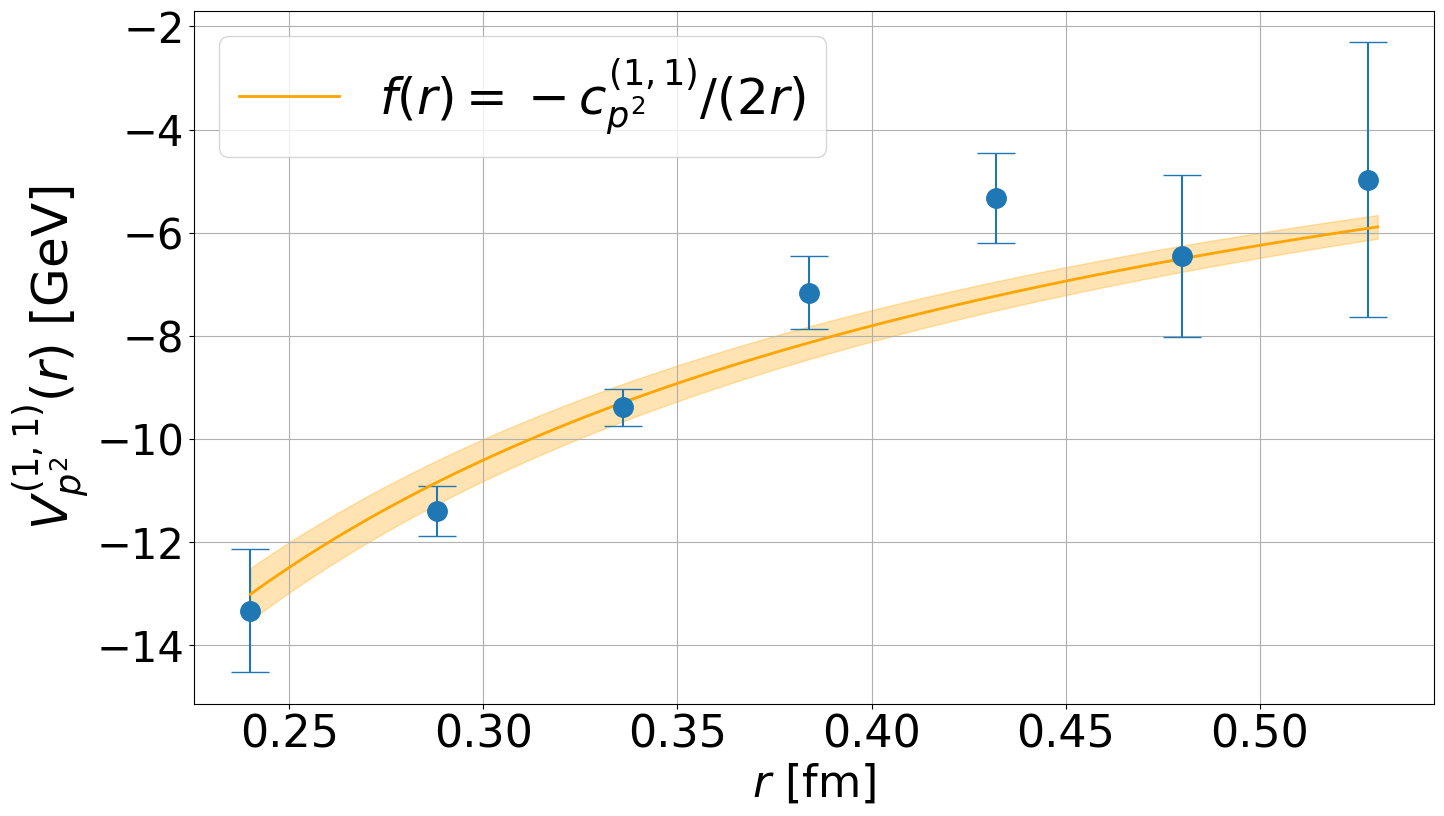}
\includegraphics[width=0.48\linewidth]{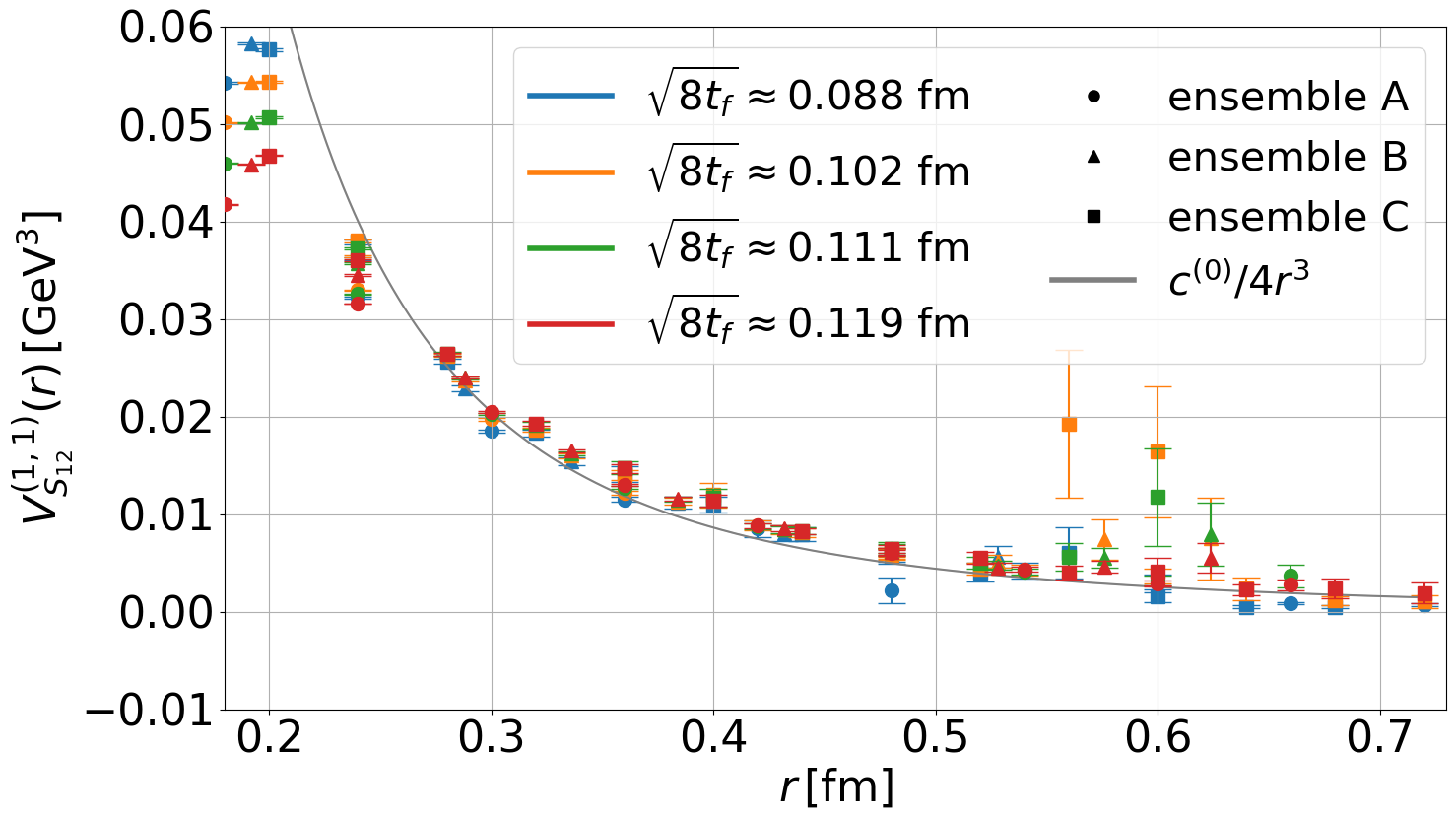}
\caption{Top left, top right and bottom left: $rV_{LS}^{(2,0)}$, $rV_{LS}^{(1,1)}$ and $V_{p^2}^{(1,1)}$ from ensemble B at $r_f = \sqrt{8t_f}\approx 0.119$ fm. Yellow lines and error bands represent fits of the ans\"atze $f(r)$ provided in the plots. Bottom right: $V_{S_{12}}^{(1,1)}$ from all ensembles and flow times. The grey curve represents the tree level result.}
\label{fig:potential_corrections}
\end{figure}

Thus, in Fig.\ \ref{fig:potential_corrections} we only present results for the potential $V_{S_{12}}^{(1,1)}$ for all ensembles and flow times, while results for the potentials $V_{LS}^{(2,0)}$, $V_{LS}^{(1,1)}$ and $V_{p^2}^{(1,1)}$ are restricted to ensemble B and $r_f = \sqrt{8 t_f} \approx 0.119$ fm.
We fit ansaetze provided in Fig.\ \ref{fig:potential_corrections} to the data points, which are motivated by the perturbative short range behavior and the long range prediction from effective string theory \cite{Barchielli:1988zp,Brambilla:2014eaa}.
We are able to determine the fit parameters quite accurately, e.g.\ the long range parameter $g\Lambda^\prime$, which has previously not been properly extracted from lattice data (for a crude estimate of $g\Lambda^\prime$ without an error analysis see Ref.\ \cite{Oncala:2017hop}). 
In the case of $V_{LS}^{(1,1)}$ (top right plot) we find that including the long range term $\sim g\Lambda^\prime$ in the fit ansatz is crucial to achieve $\chi^2/\mathrm{DOF} \sim 1$. 
As in the case of the static potential without tree level improvement, the results for $V_{S_{12}}^{(1,1)}$ (bottom right plot) at small separations $r$ are strongly affected by discretization errors and flow effects.
We also show in this plot the tree level result $C_F \alpha_s / 4 r^3 \approx c^{(0)} / 4r^3$, where $c^{(0)}$ was determined from a fit to $V^{(0)}$ data points as discussed above.
At intermediate and large $r$, where discretization errors and flow effects are expected to be small, the grey curve and the data points agree within statistical errors.

\begin{figure}[ht]
\centering
\includegraphics[width=0.48\linewidth]{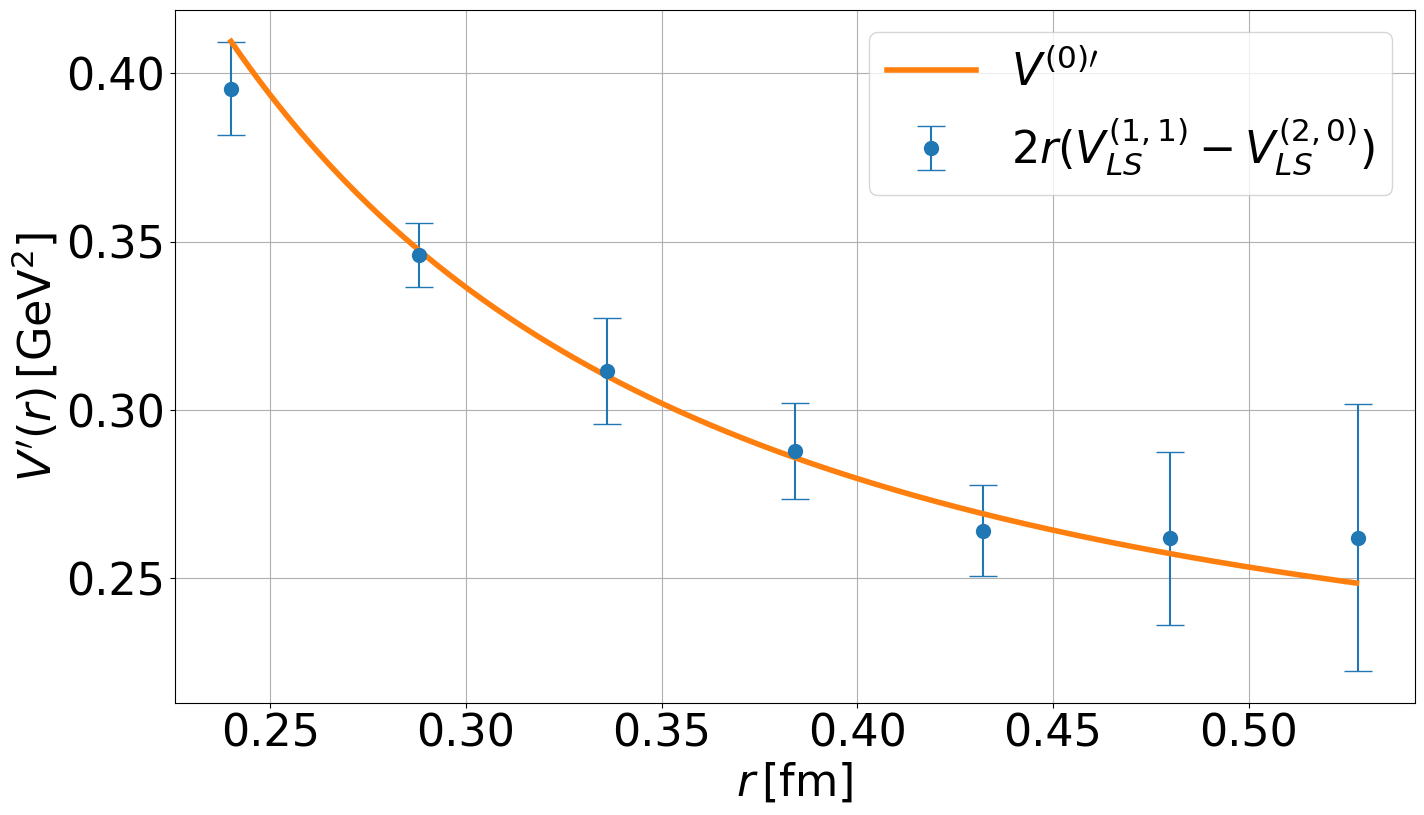}
\includegraphics[width=0.48\linewidth]{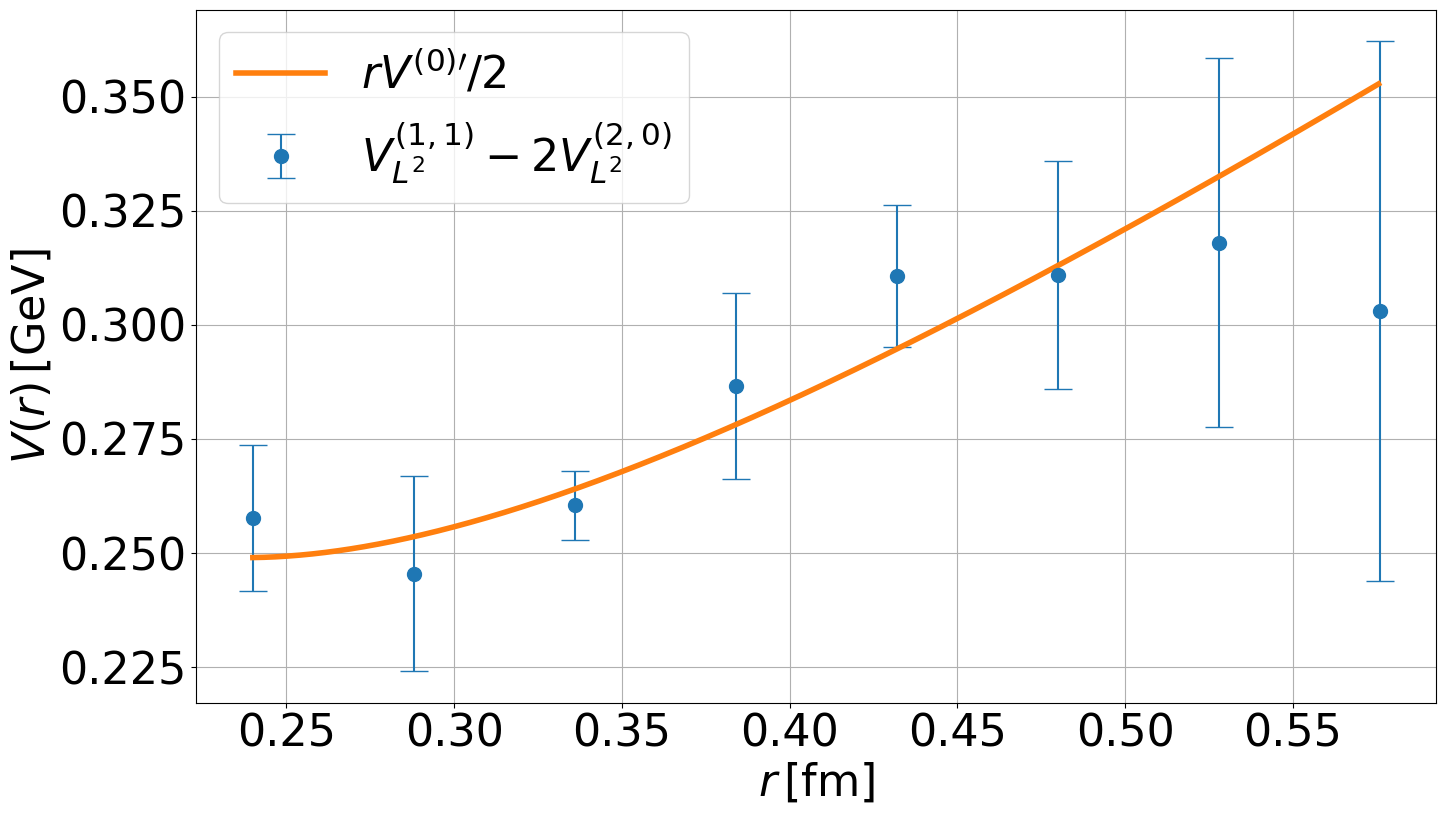}
\includegraphics[width=0.48\linewidth]{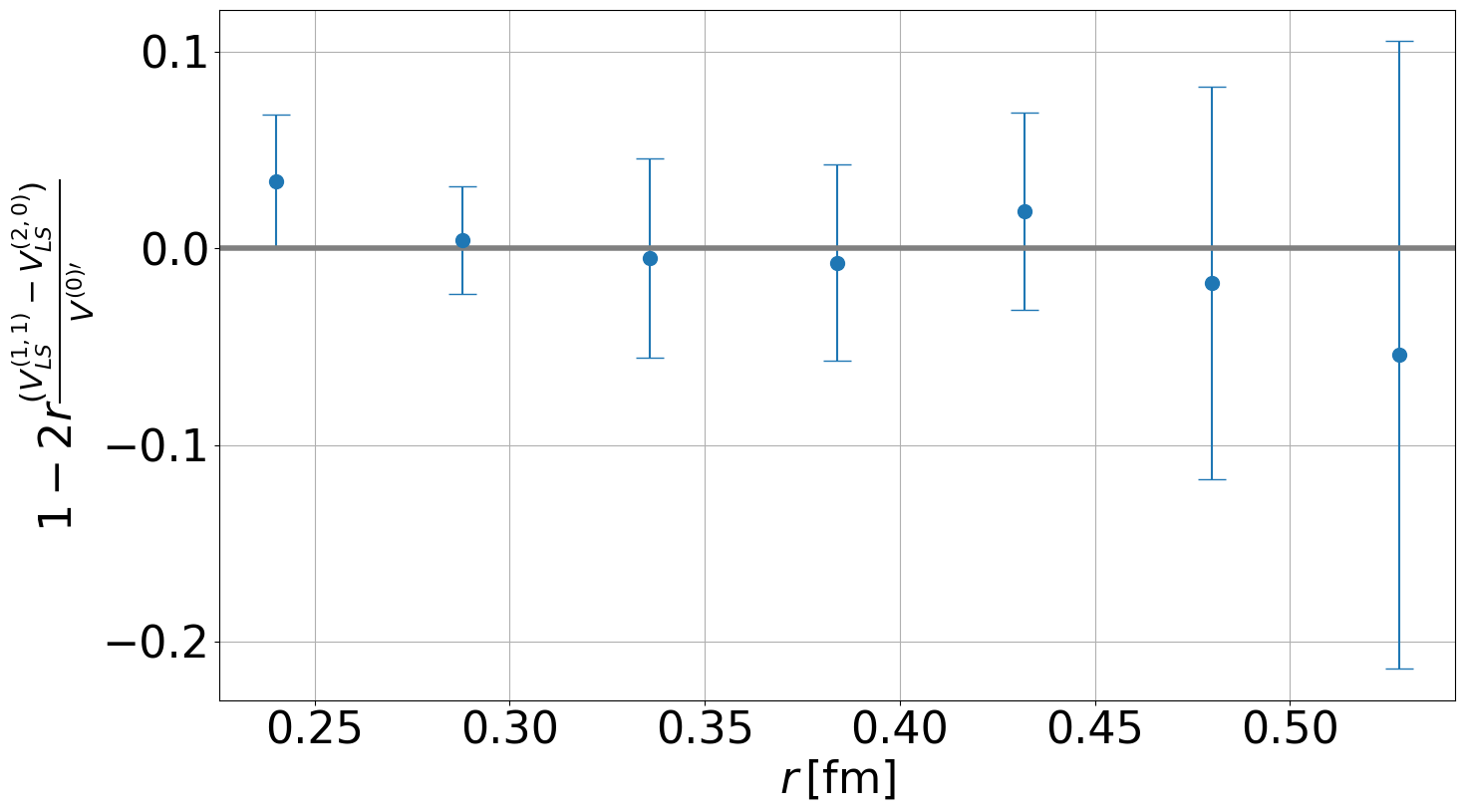}
\includegraphics[width=0.48\linewidth]{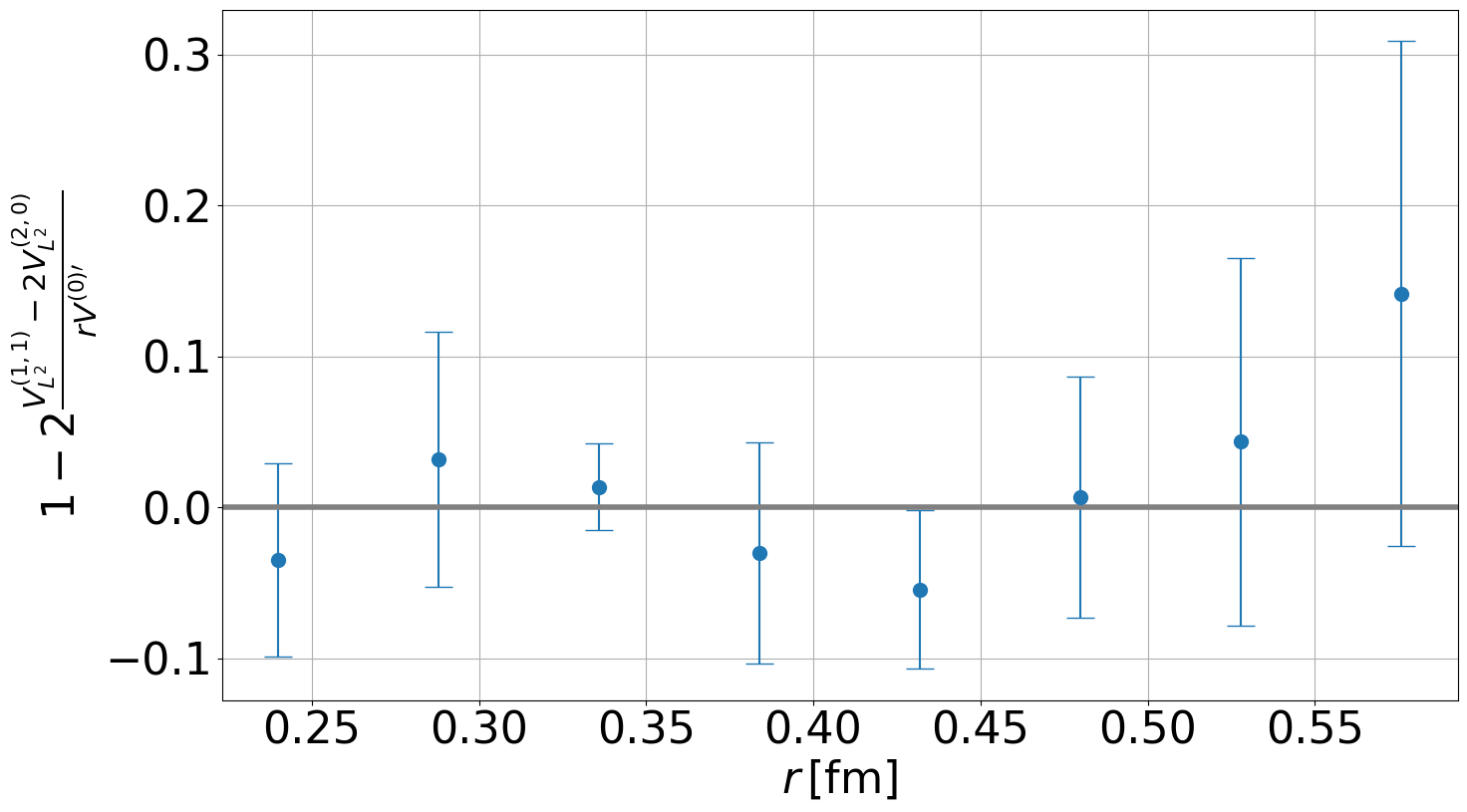}
\caption{Numerical check of the Gromes relation (\ref{eq:gromes_relation}) (left plots) and the first BBMP relation (\ref{eq:bbmp1_relation}) (right plots) for ensemble B and flow radius $r_f = \sqrt{8 t_f} \approx 0.119$ fm.}
\label{fig:gromes_bbmp}
\end{figure}

The Gromes and BBMP relations relate some of the potential corrections to $V^{(0)}$ and $V^{(0)\prime}$ are independent of NRQCD matching coefficients \cite{Gromes:1984ma, Barchielli:1988zp, Brambilla:2003nt}. 
Thus, these relations are ideally suited to study the magnitude of discretization errors and to explore, to what extent the renormalization of field strength insertions was successful.
The Gromes relation and the first BBMP relation are
\begin{align}
V_{LS}^{(1,1)} - V_{LS}^{(2,0)}
&= 
\frac{1}{2r} V^{(0)\prime}\,,
\label{eq:gromes_relation}
\\
V_{L^2}^{(1,1)} - V_{L^2}^{(2,0)} - V_{L^2}^{(0,2)}
&=
\frac{r}{2} V^{(0)\prime}\,.
\label{eq:bbmp1_relation}
\end{align}
Numerical results from ensemble B at flow radius $r_f = \sqrt{8t_f} \approx 0.119$ are shown for both relations in Fig.\ \ref{fig:gromes_bbmp}, in the top row the left hand side and the right hand side separately, in the bottom row their ratio. Even though these data points are neither continuum nor zero-flow-time extrapolated, we find that the relations are fulfilled within statistical errors.
It is interesting to note that in Ref.\ \cite{Koma:2006fw} the Gromes relation was violated by a few percent. The corresponding data points were renormalized apprroximately with the Huntley-Michael prescription \cite{Huntley:1986de}, which leaves unwanted terms of order $\mathcal{O}(g^4)$ for chromo-electric and $\mathcal{O}(g^6)$ for chromo-magnetic insertions \cite{Bali:1997am}. Gradient flow does not have such a limitation at sufficiently large $t_f > 0$. Our findings together with the findings from Ref.\ \cite{Koma:2006fw}, thus, suggest that violations of the Gromes relation in Ref.\ \cite{Koma:2006fw} were mostly caused by approximations in the Huntley-Michael renormalization prescription.

\subsection{\label{SEC_results_corrections}Tree level improvement of correlators with field strength insertions}

For the static potential tree level improvement turned out to be a successful strategy, to parameterize and to remove the majority of systematic errors due to the finite lattice spacing and flow time (see Sec.\ \ref{SEC_gradient_flow} and Sec.\ \ref{SEC_results_V0}). Now we propose and explore a similar procedure for field strength correlators.
Note that NRQCD matching coefficients, which appear in case of spin-dependent potentials, differ from $1$ at $\mathcal{O}(\alpha^2)$ and, thus, can be ignored in the following.

\begin{figure}[ht]
\centering
\includegraphics[width=0.45\linewidth]{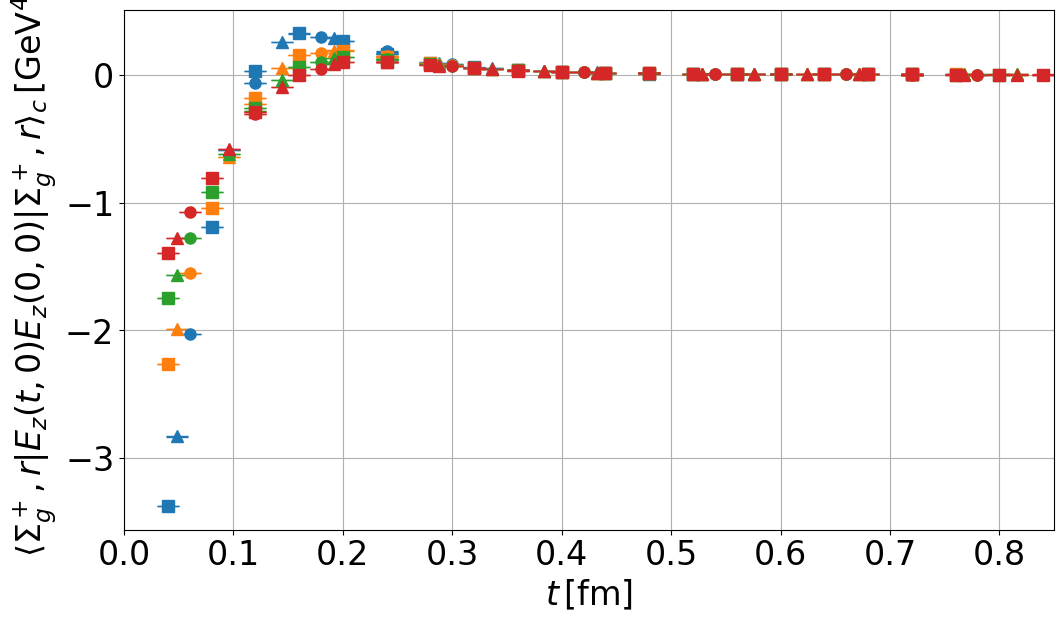}
\includegraphics[width=0.45\linewidth]{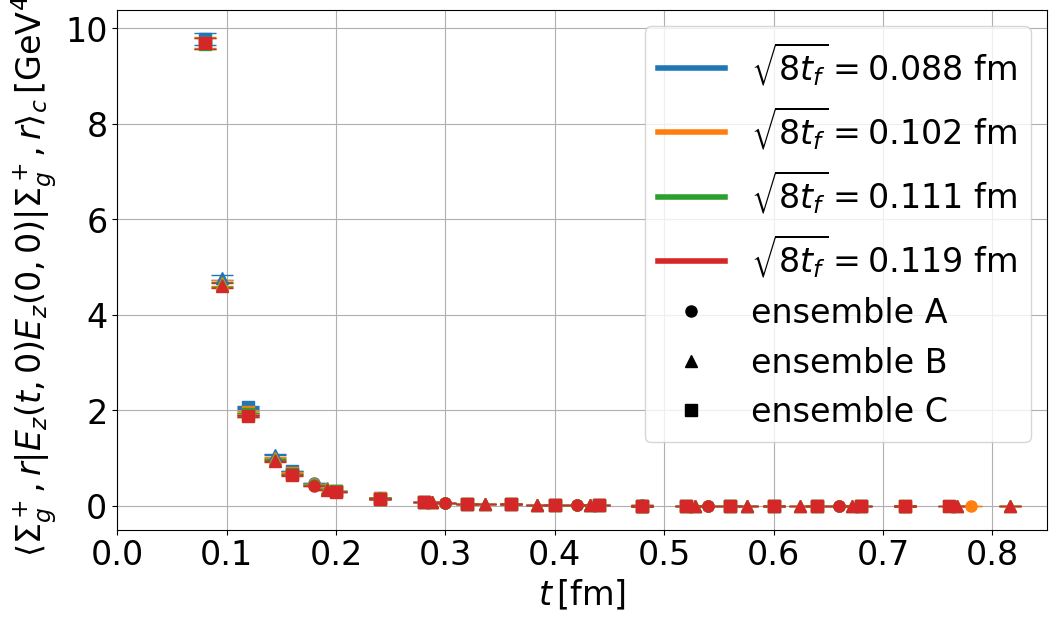}
\includegraphics[width=0.45\linewidth]{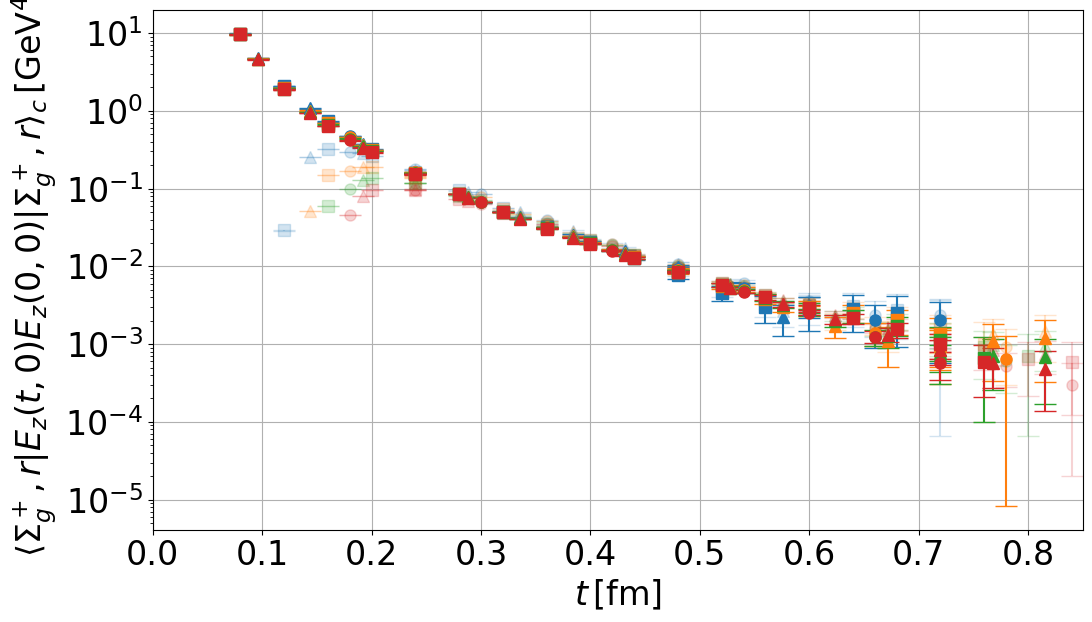}
\caption{Results for the correlator $\langle \Sigma_g^+, r \vert E_z(t, 0) E_z(0, 0) \vert \Sigma_g^+, r\rangle_c$ from all three ensembles and for several flow times for $r\approx 0.24$ fm. Top left: Unimproved data. Top right: Tree level improved data according to Eq.\ (\ref{eq:corrected_correlator}). Bottom: Unimproved data (transparent) and tree level improved data (opaque) on a logarithmic scale.}
\label{fig:improved_correlator}
\end{figure}

As an example we separate the heavy quarks along the $z$ axis und consider the field strength correlator
\begin{eqnarray}
\nonumber & & \hspace{-0.7cm} \langle \Sigma_g^+, r \vert E_z(t, 0) E_z(0, 0) \vert \Sigma_g^+, r\rangle_c = \\
\label{EQN_EzEz} & & = \langle \Sigma_g^+, r \vert E_z(t, 0) E_z(0, 0) \vert \Sigma_g^+, r\rangle - \langle \Sigma_g^+, r \vert E_z(t,0) \vert \Sigma_g^+, r\rangle \langle \Sigma_g^+, r \vert E_z(0, 0) \vert \Sigma_g^+, r\rangle \, ,
\end{eqnarray}
where the subscript $c$ denotes subtraction of the ground state expectation values of the field strength operators. This correlator is related to the potential corrections $V^{(1)}$, $V_{p^2}^{(2,0)}$ and $V_{L^2}^{(2,0)}$. Numerical results for the correlator without tree level improvement are shown for $r \approx 0.24$ fm in the top left plot of Fig.\ \ref{fig:improved_correlator}. While data points from different ensembles obtained with the same flow time $t_f$ agree quite well, there is a rather strong dependence on $t_f$ at small temporal separations $t \ltapprox r_{f,\text{max}} = \sqrt{8 t_{f,\text{max}}} \approx 0.24$ fm of the two chromoelectric operators. We recall that we are not just interested in the large-$t$ behavior of such correlators, as it is typically the case when extracting energy levels, but also need precise results at small $t$, to solve the integrals (\ref{eq:integral_type}), which provide the potential corrections.

We define tree level improved correlators as
\begin{align}
C_\text{tree level improved}^\text{lattice}(t, r, t_f)
&=
C^\text{lattice}(t, r, t_f) - \tilde{c} \bigg(\frac{4 \pi}{C_F g^2} \Big(C_\text{tree level}^\text{lattice}(t, r, t_f) - C_\text{tree level}^\text{continuum}(t, r)\Big)\bigg) \, .
\label{eq:corrected_correlator}
\end{align}
$C^\text{lattice}$ refers to the unimproved numerical results, where an example was discussed in the previous paragraph (see Fig.\ \ref{fig:improved_correlator}, top left). $C_\text{tree level}^\text{lattice}$ denotes the tree level result in lattice perturbation theory with gradient flow, which we compute following the lines of Ref.\ \cite{Fritzsch:2013je}, and $C_\text{tree level}^\text{continuum}$ is the tree level result in continuum perturbation theory without gradient flow. The constant $\tilde{c}$ can, in principle, be determined by choosing a parameterization for $C_\text{tree level improved}^\text{lattice}$ in Eq.\ (\ref{eq:corrected_correlator}) and carrying out a single global fit of $C^\text{lattice}$ to numerical data from different ensembles and for several flow times. Another simpler strategy, which led to improved correlators of similar quality, is to use $\tilde{c} = \tilde{c}^{(0)}$, where $\tilde{c}^{(0)}$ is conceptually identical to $\tilde{c}$ and was already determined, while improving the ordinary static potential (see Eq.\ (\ref{eq:ansatz_tree_level}) and Sec.\ \ref{SEC_results_V0}).

The tree level improved version of the correlator (\ref{EQN_EzEz}) is shown in Fig.\ \ref{fig:improved_correlator}, top right. The small-$t$ behavior strongly differs from its unimproved counterpart (top left plot), while there are almost no differences for large $t$. Note in particular that results obtained with different flow times now agree very well. The bottom plot in Fig.\ \ref{fig:improved_correlator} compares the improved correlator (opaque data points) and the unimproved correlator (transparent data points) on a logarithmic scale, exhibiting in a much clearer way the differences at small $t$ and the agreement at large $t$.

As a next step we plan to apply this method of tree level improvement to the full set of field strength correlators. This should provide stable and trustworthy correlator data at significantly smaller $r$ and $t$ and possibly stabilize $a\rightarrow 0$ and $t_f \rightarrow 0$ extrapolations of potential corrections.


\section*{Acknowledgements}

We acknowledge interesting and useful discussions with Gunnar Bali, Nora Brambilla, Viljami Leino, Julian Mayer-Steudte, Guy Moore, Carolin Schlosser, Antonio Vairo and Xiang-Peng Wang.
M.W.\ acknowledges support by the Heisenberg Programme of the Deutsche Forschungsgemeinschaft (DFG, German Research Foundation) – project number 399217702. 
Calculations on the GOETHE-NHR and on the on the FUCHS-CSC high-performance computers of the Frankfurt University were conducted for this research. 
We thank HPC-Hessen, funded by the State Ministry of Higher Education, Research and the Arts, for programming advice.


\end{document}